\documentclass[prl,preprint,showpacs]{revtex4}
\usepackage{graphicx}
\begin{document}
\title{Phase coherence length and quantum interference patterns at step edges}
\author{S.\ Crampin}
\affiliation{Department of Physics, University of Bath, Bath BA2 7AY, 
United Kingdom}
\author{J.\ Kr\"oger}
\author{H.\ Jensen}
\author{R.\ Berndt}
\affiliation{Institut f\"ur Experimentelle und Angewandte Physik, 
Christian-Albrechts-Universit\"at zu Kiel, D-24098 Kiel, Germany}
\maketitle

A recent Letter by Wahl \textit{et al.} \cite{wah03_} is the first report of 
measurements using the scanning tunneling microscope of the dynamical 
properties of electrons in two-dimensional image potential states. In their 
experiment the spatial decay of quantum interference patterns at step edges 
on Cu(001) is used to determine the phase coherence length $L_\Phi$ as a 
function of energy. This is converted to a linewidth 
$\Gamma=\hbar v_{\text{g}}/L_\Phi$ using the group velocity 
$v_{\text{g}}=\hbar k/m^\ast$ of the image state electrons, which have 
parabolic dispersion $E=\hbar^2 k^2/2m^\ast$, and compared to linewidths 
measured by time-resolved two-photon photoemission (2PPE) \cite{ber02_}.
However, Wahl \textit{et al.} have incorrectly identified the phase coherence 
length in the quantum interference patterns, so that the agreement between 
their linewidths and those from 2PPE must be viewed as fortuitous.

In \cite{wah03_} the local density of image electron states (LDOS) a 
distance $x$ from a step is taken to be
\begin{equation}
\rho(x)\approx L_0\left[1-|r|e^{-2x/L_\Phi} J_0(2kx)\right]
\label{eqn:wahl}
\end{equation}
where the reflection coefficient is $|r|\exp -i\pi$. In the absence of 
inelastic scattering $L_\Phi\!\rightarrow\!\infty$ and (\ref{eqn:wahl}) 
reduces to the exact result for a two-dimensional electron gas near a step. 
The distance $2x$ appears in the argument of the Bessel function $J_0$ as 
this is the path length for electrons to scatter off the step and return 
to $x$. The appearance of $2x$ in the exponential then implies that over a 
distance $x$ across the surface the image electron wave functions
decay like $\psi \sim \exp -x/L_\Phi$. The electrons take time 
$t=x/v_{\text{g}}$ to travel $x$, hence $\psi\sim \exp -v_{\text{g}} t/L_\Phi$,
 giving $|\psi|^2\sim \exp -2\Gamma t/\hbar$. But the photoemission linewidth 
is related to the quantum state lifetime $\tau$ by $\Gamma=\hbar/\tau$, and 
so one should have $|\psi|^2\sim \exp -\Gamma t/\hbar$. Hence Eqn. 
(\ref{eqn:wahl}) must contain an incorrect dependence on $L_\Phi$.

\begin{figure}[t]
\centering
\includegraphics[width=8.5cm,clip=]{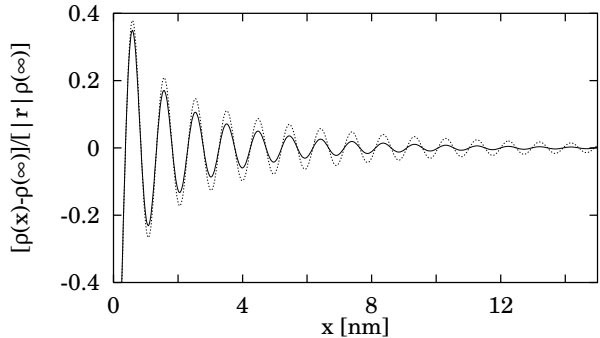}
\caption{Quantum interference patterns at a step calculated using Eqn. (1), 
the approximation used in \cite{wah03_} (solid line), and Eqn. (2), the 
exact expression (dotted line). We use image state energy 
$E=0.2$\ eV, mass $m^\ast=m_e$, linewidth $\Gamma=20$\ meV.}
\label{fig:decay}
\end{figure}

This heuristic argument is confirmed by rigorous analysis which we outline.
We calculate the LDOS by solving
$
(-\hbar^2\nabla^2/2m^\ast+V-E+i\text{Im}\,\Sigma)G=
-\delta({\bf r}-{\bf r}')
$
for the single-particle Green function $G$; inelastic particle interactions 
are included through a complex self energy 
with $-\text{Im}\,\Sigma=\hbar/(2\tau)=\Gamma/2$ \cite{kli00_,chu98_}. Then
$\rho=-(2/\pi)\text{Im}\,G({\bf r},{\bf r})$ and for $x>0$ we find
\begin{equation}
\rho(x)=\frac{m^\ast}{\pi^2\hbar^2}\text{Re}\int_{-\infty}^\infty
dq \left[1-|r|e^{2i\kappa x}\right]/\kappa
\label{eqn:exact}
\end{equation}
where $\kappa=\sqrt{\eta^2-q^2}$ and 
$\eta=\sqrt{2m^\ast (E-i\text{Im}\,\Sigma)/\hbar^2}$. 
In Fig. \ref{fig:decay} we compare the standing waves given by
(\ref{eqn:exact}) with the approximation (\ref{eqn:wahl}) used in 
Ref. \cite{wah03_}.  The latter significantly overestimates the damping of 
the quantum interference patterns.

The phase relaxation length is readily identified from (\ref{eqn:exact}) 
as follows.
Using the stationary phase approximation the asymptotic behaviour of 
$I=\int_{-\infty}^\infty dq e^{2i\kappa x}/\kappa$ can be found:
$I\!\sim\!\sqrt{\pi/\eta x}e^{2i\eta x -i\pi/4}$.\
When $|\text{Im}\,\Sigma|\ll E$ which holds in \cite{wah03_} we have 
$\eta\approx\sqrt{2m^\ast E/\hbar^2}(1-i\text{Im}\,\Sigma/2E)
= k+i/2L_\Phi$, since $k=\sqrt{2m^\ast E/\hbar^2}$ and 
$-k\text{Im}\,\Sigma/2E=\Gamma/2\hbar v_{\text{g}}$. 
Therefore $\text{Re}\,I\sim \sqrt{\pi/k x} e^{-x/L_\Phi}\cos
(2kx-\pi/4)$. Since asymptotically 
$J_0(z)\sim \sqrt{2/\pi z}\cos(z-\pi/4)$, the LDOS near a step is
correctly approximated as
\begin{equation}
\rho(x)\approx L_0\left[1-|r|e^{-x/L_\Phi} J_0(2kx)\right],
\label{eqn:approx}
\end{equation}
which has the form used in \cite{wah03_} but with $L_\Phi$ replaced 
by $2L_\Phi$.  The phase relaxation lengths extracted using (\ref{eqn:wahl}) 
should be halved, and hence the linewidths in \cite{wah03_} doubled.

\smallskip
S. C. acknowledges the support of the British Council. J. K., H. J, and 
R. B. thank the Deutsche Forschungsgemeinschaft for financial support.

\end{document}